\documentclass[aps,preprint]{revtex4}%
\usepackage{amsfonts}
\usepackage{amsmath}
\usepackage{amssymb}
\usepackage{graphicx}%
\begin{document}
\title[ ]{Derivation of the \ Planck Spectrum for Relativistic Classical Scalar
Radiation from Thermal Equilibrium in an Accelerating Frame}
\author{Timothy H. Boyer}
\affiliation{Department of Physics, City College of the City University of New York, New
York, New York 10031}

\begin{abstract}
The Planck spectrum of thermal scalar radiation is derived suggestively within
classical physics by the use of an accelerating coordinate frame. \ The
derivation has an analogue in Boltzmann's derivation of the Maxwell velocity
distribution for thermal particle velocities by considering the thermal
equilibrium of noninteracting particles in a uniform gravitational field.
\ For the case of radiation, the gravitational field is provided by the
acceleration of a Rindler frame through Minkowski spacetime. \ Classical
zero-point radiation and relativistic physics enter in an essential way in the
derivation which is based upon the behavior of free radiation fields and the
assumption that the field correlation functions contain but a single
correlation time in thermal equilibrium. \ The work has connections with the
thermal effects of acceleration found in relativistic quantum field theory.

\end{abstract}
\maketitle

\section{Introduction}

Many text books present Boltzmann's derivation\cite{Boltzmann} of the Maxwell
velocity distribution for free particles in thermal equilibrium in a box. In
his analysis, Boltzmann introduced a uniform gravitational field, followed the
implications of thermal equilibrium under gravity, and finally took the
zero-gravity limit. \ The derivation is striking because it uses only the
physics of free nonrelativistic particles moving in a gravitational field.
\ By the principle of equivalence, the gravitational field can be replaced by
an accelerating coordinate frame. \ But then thermodynamic consistency
requires that the interactions of particles which lead to equilibrium in an
inertial frame must be consistent with the equilibrium determined by the
physics of free particles in an accelerating frame. \ The natural question
arises as to whether the analogue of this procedure can be applied to the much
more complicated problem of thermal equilibrium for relativistic radiation
with its infinite number of normal modes. \ In this article we show that an
analogous derivation is indeed possible for relativistic classical scalar
radiation. \ We introduce a relativistic accelerating coordinate frame (a
Rindler frame, which is the closest relativistic equivalent to a uniform
gravitational field), consider the implications for thermal radiation
equilibrium, make the assumption that thermal equilibrium involves but a
single correlation time, and finally take the limit of zero acceleration to
obtain the thermal radiation spectrum in an inertial frame. \ The use of an
accelerating coordinate frame to obtain the thermal equilibrium spectrum seems
striking because only noninteracting free radiation fields are needed for the
derivation. \ However, we expect that any other interaction which produces
equilibrium must be consistent with the equilibrium determined by the
accelerating frame. \ In particular, the use of nonrelativistic nonlinear
scattering systems\cite{nonlinear} which lead to the Rayleigh-Jeans spectrum
for radiation equilibrium are inconsistent with special relativity and
relativistic accelerating coordinate frames.

The derivation here for the Planck spectrum is provided within the context of
relativistic classical scalar field theory which includes classical zero-point
radiation. \ This classical \textit{scalar} field theory is analogous to the
classical \textit{electromagnetic} theory with classical electromagnetic
zero-point radiation. \ The classical electromagnetic theory has been shown to
provide classical explanations for a number of phenomena which are usually
regarded as belonging to the exclusive domain of quantum theory, including
Casimir forces, van der Waals forces, diamagnetism, specific heats of
solids,\cite{review}\cite{delaP} and the ground state of hydrogen.\cite{Cole}
\ The description of thermal radiation in terms of classical radiation with
random phases which is used in the classical theories is a standard procedure
dating from nineteenth century physics. \ The choice of zero-point radiation
as the "vacuum" situation in classical physics is required in order to
describe the experimentally observed Casimir forces, but this choice then
gives natural classical explanations for other phenomena. \ The Lorentz
invariance of classical zero-point radiation determines the spectrum up to one
unknown multiplicative constant giving the scale of the zero-point radiation.
\ The scale of zero-point radiation is chosen to give numerical agreement with
experimental measurements of Casimir forces. \ It turns out that the unknown
multiplicative constant takes a numerical value which is immediately
recognizable as Planck's constant $\hbar.$ \ Thus Planck's constant $\hbar$
enters the classical theory as the scale factor of classical zero-point
radiation and not as the quantum of action familiar in current quantum theory.

There have been many indignant objections to work involving "classical"
zero-point radiation; the claim is made that zero-point radiation is
exclusively a "quantum" concept. \ The classical electromagnetic theory
treated earlier and the classical scalar theory discussed here are both
"classical" in the sense that they contain no intrinsically discrete aspects
of energy or action. \ The zero-point radiation is classical random radiation
chosen as a perfectly valid homogeneous boundary condition on the classical
field equations. \ The concept of zero-point radiation (random radiation
fluctuations at the zero of temperature) can appear in both classical and
quantum theories. \ Zero-point radiation can not be regarded as belonging
exclusively to quantum theory any more than the concepts of mass, energy, and
gravity can be claimed as exclusively classical concepts because they appeared
first in the context of classical mechanics.

The outline of our presentation is as follows. \ In Section II, we review the
basics of classical relativistic scalar field theory. We introduce the random
phases between normal modes for stationary distributions of random classical
radiation, and then we calculate the two-point field correlation function
associated with a general stationary spectrum of random radiation. In Section
III, we discuss thermal radiation in an inertial frame. \ We start with the
two fundamental ideas required for understanding thermal radiation equilibrium
within classical physics. \ These include the presence of a divergent spectrum
of classical zero-point radiation and the presence of a finite density of
thermal radiation above the zero-point spectrum. \ We note that thermodynamic
ideas give us Wien's displacement law and the Stefan-Boltmann relation.
Although Wien's law gives a restriction on the form of the radiation spectrum
and also on the form of the two-point field correlation function,
thermodynamics in an inertial frame does not determine the spectrum of thermal
radiation. \ \ In Section IV we introduce a relativistic coordinate frame (a
Rindler frame) accelerating through Minkowski spacetime. We note the role
played by acceleration in breaking the homogeneity and isotropy of Minkowski
spacetime. \ Then we review some preliminary information regarding the Rindler
accelerated coordinate frame, and use the thermodynamics of pressure
equilibrium to show that temperature and acceleration have the same spatial
dependence throughout a Rindler frame. Next we recalculate the two-point
correlation function for classical zero-point radiation in terms of Rindler
coordinates. \ But then one sees a natural behavior for the thermal
correlation function which follows from the known correlation function
involving zero-point radiation. \ Finally we take the limit of vanishing
acceleration and so recover the Planck spectrum as the classical radiation
spectrum of thermal equilibrium. \ The article closes with allusions to
related but vastly different work in relativistic quantum field theory.

\section{Scalar Field Theory for Random Fields}

\subsection{The Relativistic Scalar Field}

In an inertial frame, the relativistic free scalar field $\phi(ct,x,y,z)$ is
specified by the Lagrangian density\cite{Goldstein}%
\begin{equation}
\mathcal{L}=\frac{1}{8\pi}\left[  \frac{1}{c^{2}}\left(  \frac{\partial\phi
}{\partial t}\right)  ^{2}-\left(  \frac{\partial\phi}{\partial x}\right)
^{2}-\left(  \frac{\partial\phi}{\partial y}\right)  ^{2}-\left(
\frac{\partial\phi}{\partial z}\right)  ^{2}\right]
\end{equation}
which leads to the wave equation as the equation of motion%
\begin{equation}
\frac{1}{c^{2}}\frac{\partial^{2}\phi}{\partial t^{2}}\mathbf{-}\frac
{\partial^{2}\phi}{\partial x^{2}}\mathbf{-}\frac{\partial^{2}\phi}{\partial
y^{2}}\mathbf{-}\frac{\partial^{2}\phi}{\partial z^{2}}=0
\end{equation}
The energy $U$ in the field follows from the Lagrangian density in (1)
as\cite{Goldstein}%
\begin{equation}
U=\int d^{3}x\frac{1}{8\pi}\left[  \frac{1}{c^{2}}\left(  \frac{\partial\phi
}{\partial t}\right)  ^{2}+\left(  \frac{\partial\phi}{\partial x}\right)
^{2}+\left(  \frac{\partial\phi}{\partial y}\right)  ^{2}+\left(
\frac{\partial\phi}{\partial z}\right)  ^{2}\right]
\end{equation}

The radiation in a box can be described by a complete set of either standing
waves or running waves with appropriate wave vectors $\mathbf{k.}$ $\ $In the
present case, we are not interested in any special conditions holding at the
walls of the rectangular box of dimensions $L_{x}\times L_{y}\times L_{z},$
and so we will choose periodic running waves where
\begin{equation}
\mathbf{k}=\widehat{x}(n_{x}2\pi/L_{x})+\widehat{y}(n_{y}2\pi/L_{y}%
)+\widehat{z}(n_{z}2\pi/L_{z})
\end{equation}
and the integers $n_{x},n_{y},n_{z}$ run over all positive and negative
values. \ Then the radiation field in the box can be written as
\begin{equation}
\phi(ct,x,y,z)=\sum_{n_{x}=-\infty}^{\infty}\sum_{n_{y}=-\infty}^{\infty}%
\sum_{n_{z}=-\infty}^{\infty}\frac{f(c\mathbf{k)}}{(L_{x}L_{y}L_{z})^{1/2}%
}\cos[\mathbf{k\cdot r}-kct-\theta(\mathbf{k})]
\end{equation}
where $k=\left\vert \mathbf{k}\right\vert ,$ and $\theta(\mathbf{k})$ is an
appropriate phase. \ Each mode $\phi_{\mathbf{k}}(ct,x,y,z)=[f(c\mathbf{k)/}%
(L_{x}L_{y}L_{z})^{1/2}]\cos[\mathbf{k\cdot r}-kct-\theta(\mathbf{k})]$
labeled by $n_{x},n_{y},n_{z}$ has the energy $U_{\mathbf{k}}$ found by
substituting into equation (3),%
\begin{equation}
U_{\mathbf{k}}=\frac{1}{8\pi}k^{2}f^{2}(c\mathbf{k})
\end{equation}

\subsection{Two-Point Correlation Function for Random Radiation}

Coherent radiation involves fixed phase relations $\theta(\mathbf{k}%
)-\theta(\mathbf{k}^{\prime})$ between the various modes $\phi_{\mathbf{k}}$
and $\phi_{\mathbf{k}^{\prime}}$ which are used to decompose a radiation
pattern. \ Random radiation involves the opposite situation. \ Random
radiation can be written in the form of Eq. (5) where the phases
$\theta(\mathbf{k})$ are randomly distributed on the interval $[0,2\pi)$ and
are independently distributed for each $\mathbf{k.}$ \ It is convenient to
characterize random radiation by taking the two-point correlation function of
the fields $\left\langle \phi(ct,x,y,z)\phi(ct^{\prime},x^{\prime},y^{\prime
}z^{\prime})\right\rangle $ obtained by averaging over the random phases as%
\begin{equation}
\left\langle \cos\theta(\mathbf{k)}\cos\theta\mathbf{(k}^{\prime
})\right\rangle =\left\langle \sin\theta(\mathbf{k)}\sin\theta\mathbf{(k}%
^{\prime})\right\rangle =(1/2)\delta_{\mathbf{k,k}^{\prime}}%
\end{equation}%
\begin{equation}
\left\langle \cos\theta(\mathbf{k)\sin\theta(k}^{\prime})\right\rangle =0
\end{equation}
\ Then the two-point correlation function for a general isotropic distribution
of classical scalar waves is\cite{corr}%
\begin{align}
\left\langle \phi(ct,x,y,z)\phi(ct^{\prime},x^{\prime},y^{\prime}z^{\prime
})\right\rangle  &  =<\sum_{n_{x}=-\infty}^{\infty}\sum_{n_{y}=-\infty
}^{\infty}\sum_{n_{z}=-\infty}^{\infty}\frac{f(c\mathbf{k)}}{(L_{x}L_{y}%
L_{z})^{1/2}}\cos[\mathbf{k\cdot r}-kct-\theta(\mathbf{k})]\nonumber\\
&  \times\sum_{n_{x}^{^{\prime}}=-\infty}^{\infty}\sum_{n_{y}^{^{\prime}%
}=-\infty}^{\infty}\sum_{n_{z}^{^{\prime}}=-\infty}^{\infty}\frac
{f(c\mathbf{k}^{^{\prime}}\mathbf{)}}{(L_{x}L_{y}L_{z})^{1/2}}\cos
[\mathbf{k}^{^{\prime}}\mathbf{\cdot r}^{^{\prime}}-k^{^{\prime}}ct^{^{\prime
}}-\theta(\mathbf{k}^{^{\prime}})]>\nonumber\\
&  =\frac{1}{2}\sum_{n_{x}=-\infty}^{\infty}\sum_{n_{y}=-\infty}^{\infty}%
\sum_{n_{z}=-\infty}^{\infty}\frac{f^{2}(c\mathbf{k)}}{L_{x}L_{y}L_{z}}%
\cos[\mathbf{k\cdot(r-r}^{\prime})-kc(t-t^{\prime})]
\end{align}
For a very large box, the normal modes are closely spaced and the sums over
the integers $n_{x},n_{y},n_{z}$ can be replaced by integrals of the form
$dk_{x}=(2\pi/L_{x})dn_{x},dk_{y}=(2\pi/L_{y})dn_{y},$ $dk_{y}=(2\pi
/L_{y})dn_{y}$ so that the correlation function of Eq. (9) becomes
\begin{equation}
\left\langle \phi(ct,x,y,z)\phi(ct^{\prime},x^{\prime},y^{\prime}z^{\prime
})\right\rangle =\frac{1}{16\pi^{3}}\int d^{3}kf^{2}(c\mathbf{k}%
)\cos[\mathbf{k\cdot(r-r}^{\prime})-\omega(t-t^{\prime})]
\end{equation}

In the work to follow, we will be interested in isotropic distributions of
random radiation so that the function $f(c\mathbf{k)}$ involves only the
magnitude $ck=\omega.$ \ In this case, we can carry out the angular
integrations for $\mathbf{k}$ in Eq. (10)\cite{corr}%
\begin{align}
\left\langle \phi(ct,x,y,z)\phi(ct^{\prime},x^{\prime},y^{\prime}z^{\prime
})\right\rangle  &  =\frac{1}{16\pi^{3}}\int d^{3}kf^{2}(ck)\cos
[\mathbf{k\cdot(r-r}^{\prime})-\omega(t-t^{\prime})]\nonumber\\
&  =\frac{1}{8\pi^{2}c^{2}\left\vert \mathbf{r-r}^{\prime}\right\vert }%
\int_{0}^{\infty}d\omega\,\omega\,f^{2}(\omega)\{\sin[(\omega/c)(\left\vert
\mathbf{r-r}^{\prime}\right\vert -c(t-t^{\prime}))]\nonumber\\
&  +\sin[(\omega/c)(\left\vert \mathbf{r-r}^{\prime}\right\vert +c(t-t^{\prime
}))]\}
\end{align}
This integral expression (11) is as far as we can carry the evaluation of the
two-field correlation function for random radiation without knowing something
(beyond isotropy) about the spectral function $f^{2}(\omega).$

\section{Thermal Radiation in an Inertial Frame}

\subsection{Two Fundamental Ideas of Classical Thermal Radiation}

In order to understand thermal radiation within classical physics, two
fundamental ideas are needed. \ The first needed idea is the presence of
classical zero-point radiation as the universal homogeneous boundary condition
on radiation equations. \ This random zero-point radiation corresponds to the
"vacuum state" of classical physics.\cite{review} \ Zero-point radiation
exists only for massless waves within the context of relativistic theory.
\ This wave situation is in contrast with nonrelativistic physics where all
momentum is carried by mass. Within relativistic classical physics, zero-point
radiation is random classical radiation with a Lorentz-invariant spectrum.
Thus the same zero-point spectrum appears in every inertial frame. \ The
requirement of Lorentz-invariance fixes the spectrum of random zero-point
radiation up to one over-all multiplicative constant.\cite{corr}%
\cite{Marshall}\cite{boyer68} \ The second needed idea is that thermal
radiation with $T>0$ represents a finite density of random radiation above the
divergent spectrum of zero-point radiation. \ It is the divergent spectrum of
zero-point radiation which prevents the finite energy density of thermal
radiation from leaking out to the infinite number of high frequency modes.
\ For each normal mode, the thermal energy is added on top of the zero-point
energy which is always present. \ Now the idea of a finite energy density on
top of a divergent energy density may give one pause. \ \ However, any
particle system with mass interacts with only low-frequency modes; the
very-high frequency modes act too rapidly to influence a massive system.
\ Thus for example, a box with real conducting walls becomes transparent to
very high-frequency electromagnetic waves; radiowaves are reflected by a
copper sheet while gamma rays easily pass through. \ Thermal radiation can be
confined in a box whose walls reflect the low-frequency waves while the high
frequency modes carrying only zero-point energy penetrate through the walls.
\ Indeed, since the zero-point spectrum is invariant under an adiabatic
compression, the thermal radiation can be compressed while the zero-point
spectrum remains unchanged. \ Thermodynamic equilibrium involves the
distribution of energy among a very large number of weakly interacting
systems. \ In the case of radiation, each normal mode of oscillation in the
confining box can be regarded as a separate system, and the problem of
classical thermal radiation is to determine the equilibrium distribution of
energy among the modes of the box. \ 

\subsection{Thermodynamics of a Normal Mode}

Each normal mode of oscillation for the radiation in the box acts as a
harmonic oscillator system.\cite{Power} \ In a discussion of the
thermodynamics of harmonic oscillator systems in an \ inertial
frame,\cite{ThermoSHO} it was shown that thermodynamics alone requires that
the energy $U(\omega,T)$ per normal mode at frequency $\omega$ must take the
form%
\begin{equation}
U(\omega,T)=\omega F(T/\omega)
\end{equation}
where $T$ is the temperature of the system, and $F$ is some unknown function.
\ This result corresponds to Wien's displacement theorem. \ In the limit as
the temperature goes to zero $T\rightarrow0,$ the energy of a normal mode
becomes independent of $T$ and reduces to the zero-point value
\begin{equation}
U(\omega,0)=(1/2)\hbar\omega
\end{equation}
(corresponding to $F(T/\omega)\rightarrow\hbar/2)$ where the functional
dependence $U(\omega,0)=const\times\omega$ agrees with Lorentz invariance, and
the scale $\hbar/2$ is determined so as to give agreement with Casimir forces.
\ The thermal energy $U_{T}$ of the mode is the energy above the zero-point
energy and is just the difference%
\begin{equation}
U_{T}(\omega,T)=U(\omega,T)-U(\omega,0)
\end{equation}
In the limit as the temperature $T$ becomes large, the mode energy becomes
independent of the frequency $\omega$%
\begin{equation}
U(\omega,T)\rightarrow k_{B}T
\end{equation}
giving the Rayleigh-Jeans spectrum of radiation (corresponding to
$F(T/\omega)\rightarrow T/\omega)$. \ The general thermal radiation spectrum
$U(\omega,T)=\omega F(T/\omega)$ must interpolate between these two limits
where the ratio $T/\omega$ goes to zero or to infinity. \ In the earlier
discussion\cite{ThermoSHO} of the thermodynamics of a harmonic oscillator, it
was noted that the smoothest interpolation mathematically between the entropy
of these limits was that corresponding to the Planck radiation spectrum
including zero-point energy. \ Here, rather than using mathematical
considerations, we wish to use physical ideas involving accelerating frames to
obtain the spectral function connecting the high- and low-temperature limits.

\subsection{Thermal Radiation in a Box}

Thermal radiation involves a finite energy density above the zero-point
radiation spectrum. \ Thus in a container of volume $V$ with radiation modes,
the total thermal energy $\mathcal{U}(T)$ is a sum over all normal modes of
the thermal energy $U_{T}(\omega,T)$ in each mode, thermal energy being energy
above the zero-point energy as in (14)
\begin{equation}
\mathcal{U}(T)=\sum_{\omega}U_{T}(\omega,T)=\sum_{\omega}[U(\omega
,T)-U(\omega,0)]
\end{equation}
The thermal radiation will be isotropic in the inertial frame where the box is
at rest. \ For finite temperature $T>0\,$, radiation can be "thermal" in only
one coordinate frame in which its spectrum is isotropic; any observer moving
with finite velocity relative to this frame detects a spectrum which is not
isotropic. \ On the other hand, the (divergent) Lorentz-invariant zero-point
radiation is isotropic in every inertial frame. \ 

The number of normal modes per unit volume per unit frequency interval
is\cite{modes}
\begin{equation}
d\mathcal{N}=\omega^{2}d\omega/(2\pi^{2}c^{3})
\end{equation}
which, except for a factor of two, is the same as the familiar electromagnetic
case. \ The thermal energy density $u=\mathcal{U}(T)/V$ is then
\begin{align}
u(T)  &  =\int d\mathcal{N}U_{T}(\omega,T)=\int_{0}^{\infty}d\omega
\frac{\omega^{2}}{2\pi^{2}c^{3}}\omega\lbrack F(T/\omega)-F(0)]\nonumber\\
&  =T^{4}\int_{0}^{\infty}dz\frac{z^{2}}{2\pi^{2}c^{3}}z[F(1/z)-F(0)]=a_{Ss}%
T^{4}%
\end{align}
where $U_{T}(\omega,T)$ is the thermal energy (above the zero-point energy) in
a mode of frequency $\omega,$ the function $F$ is the unknown function
appearing in the thermal radiation spectrum of Eq. (12), and $a_{Ss}$ is a
constant playing the same role as Stefan's constant,\cite{Stefan} but now for
the scalar radiation field. \ \ Also, we expect the pressure $p$ to be a
function of temperature alone and to satisfy
\begin{equation}
p(T)=(1/3)u(T)
\end{equation}
where the factor of $1/3$ arises from the isotropic angular dependence of the
radiation. \ 

\subsection{Field Correlation for Thermal Radiation}

\ In the case of thermal radiation, we know from Eqs. (6) and (12)\ that the
spectral function $f_{T}(\omega)$ for the radiation field takes the form
$f_{T}^{2}(\omega)=8\pi c^{2}U_{\omega}/\omega^{2}=8\pi c^{2}F(\omega
/T)/\omega$ so that the two-point correlation function in Eq. (11) becomes%
\begin{align}
\left\langle \phi_{T}(ct,x,y,z)\phi_{T}(ct^{\prime},x^{\prime},y^{\prime
}z^{\prime})\right\rangle  &  =\frac{1}{\pi\left\vert \mathbf{r-r}^{\prime
}\right\vert }\int_{0}^{\infty}d\omega\,\,F(\omega/T)\{\sin[(\omega
/c)(\left\vert \mathbf{r-r}^{\prime}\right\vert -c(t-t^{\prime})]\nonumber\\
&  +\sin[(\omega/c)(\left\vert \mathbf{r-r}^{\prime}\right\vert +c(t-t^{\prime
})]\}\nonumber\\
&  =T^{2}\frac{1}{\pi(T\left\vert \mathbf{r-r}^{\prime}\right\vert )}\int
_{0}^{\infty}dv\,\,F(v)\{\sin[(v/c)(T\left\vert \mathbf{r-r}^{\prime
}\right\vert -Tc(t-t^{\prime})]\nonumber\\
&  +\sin[(v/c)(T\left\vert \mathbf{r-r}^{\prime}\right\vert +Tc(t-t^{\prime
})]\}
\end{align}
We can also specialize the situation to the case were the spatial separation
becomes small $\left\vert \mathbf{r-r}^{\prime}\right\vert \rightarrow0.$ \ In
this limit, the field correlation function at a single spatial coordinate
point ($x,y,z)=$ ($x^{\prime},y^{\prime},z^{\prime})$ but at two different
times $t$ and $t^{\prime}$ becomes%
\begin{align}
\left\langle \phi_{T}(ct,x,y,z)\phi_{T}(ct^{\prime},x,y,z)\right\rangle  &
=T^{2}\frac{2}{\pi c}\int_{0}^{\infty}dv\,v\,F\,(v)\cos[vT(t-t^{\prime
})]\nonumber\\
&  =T^{2}\mathfrak{F}[T(t-t^{\prime})]
\end{align}
where $\mathfrak{F}[T(t-t^{\prime})]$ is some unknown function of temperature
multiplied by time. \ 

\subsection{Zero-Point Radiation Correlation Function in an Inertial Frame}

In our discussion so far, we do not know the spectral function $f_{T}%
^{2}(\omega)=8\pi c^{2}U(\omega,T)/\omega^{2}=8\pi c^{2}F(\omega/T)/\omega$
for thermal radiation at non-zero temperature. \ However, we have stated the
spectral form for zero-point radiation in Eq. (13) based upon the
Lorentz-invariance of the spectrum. \ Substituting the expression of Eq. (13)
into Eq. (11), we find that the two-point field correlation function of the
zero-point radiation field $\phi_{0}(ct,x,y,z)$ can be calculated in closed
form as\cite{corr1}%
\begin{equation}
\left\langle \phi_{0}(ct,x,y,z)\phi_{0}(ct^{\prime},x^{\prime},y^{\prime
}z^{\prime})\right\rangle =\frac{-\hbar c}{\pi\lbrack c^{2}(t-t^{\prime}%
)^{2}-(x-x^{\prime})^{2}-(y-y^{\prime})^{2}-(z-z^{\prime})^{2}]}%
\end{equation}
The subscript $0$ on the fields on the left-hand side indicates that
zero-point radiation is involved. \ The denominator on the right-hand side
involves exactly the Lorentz-invariant square of the spacetime interval
between the two coordinate points and shows clearly the Lorentz-invariant
character of the random zero-point radiation.\ The denominator corresponds to
the square of the proper time interval between the two points as measured in
any inertial frame.

If the spatial coordinates $x,y,z$ are the same for the two points, then the
correlation function in (22) becomes%
\begin{equation}
\left\langle \phi_{0}(ct,x,y,z)\phi_{0}(ct^{\prime},x,y,z)\right\rangle
=\frac{-\hbar c}{\pi c^{2}(t-t^{\prime})^{2}}%
\end{equation}
We notice that this form is consistent with the thermal expression in Eq. (21)
provided $\mathfrak{F(}z)$ goes as the inverse of its argument squared at
small arguments $\mathfrak{F(}z)\sim-\hbar/(\pi cz^{2})$%
\begin{equation}
T^{2}\mathfrak{F}[T(t-t^{\prime})]\approx T^{2}\frac{-\hbar}{\pi
c[T(t-t^{\prime})]^{2}}=-\frac{\hbar c}{\pi c^{2}(t-t^{\prime})^{2}}%
\end{equation}
Small arguments for the function $\mathfrak{F}[T(t-t^{\prime})]$ in Eq. (21)
can refer to either small temperatures at finite time differences or small
time differences at finite temperatures. \ Agreement of the correlation
functions in Eqs. (23) and (24) shows that we expect that at large frequencies
(corresponding to short correlation times) the spectral function goes over to
the zero-point spectrum for any temperature. \ Indeed this is what we expect
when we think of the thermal radiation as being distributed among only the
lower frequency modes. \ 

The correlation function (23) for the zero-point fields involves simply the
time difference between the two spacetime points in the inertial frame without
any characteristic correlation time appearing in the expression. \ In
contrast, we expect that thermal radiation will indeed involve a correlation
time associated with the finite density of thermal radiation. \ Unfortunately,
the correlation for zero-point radiation given in Eq. (23) gives us no hint
about the low-frequency (long-time-correlation) behavior of the thermal
radiation spectrum.

\section{Use of Acceleration to Derive the Thermal Distribution}

\subsection{Review of Boltzmann's Derivation for the Maxwell Velocity
Distribution}

The use of mechanical and thermodynamic ideas in an inertial frame allows one
to obtain significant information about the thermal equilibrium distributions
of particles or of waves. \ Thus momentum transfer to the walls of a container
relates the pressure $p$\ of a gas of free particles or of radiation to the
energy density $u$ at the walls; $p=(2/3)u$ for free particles and $p=(1/3)u$
for radiation. \ The equations of state, $pV=Nk_{B}T$ for free particles and
the assumption that the energy density $u$ is a function of temperature $T$
alone for radiation, when combined with thermodynamic ideas, allow
determinations of the entropy of free particles and the energy density and
entropy for radiation. \ Indeed Wein's displacement theorem (given here in Eq.
(12) from the thermodynamics of each normal mode) is consistent with the
Stefan-Boltzmann law $u=a_{Ss}T^{4}$ appearing here in Eq. (18).

Although these mechanical and thermodynamic arguments give us considerable
information about the thermodynamics of free particles and of radiation, these
arguments do not tell us the thermal distribution of particle velocities or
the spectrum of thermal radiation in an inertial frame. \ We expect the
thermal distributions to be homogeneous in space and isotropic in direction;
however, the distribution in energy does not follow from symmetries under
space translation and rotation. The presence of gravity or acceleration breaks
the symmetry of the space and so allows one to distinguish thermal systems.
\ For free nonrelativistic particles, this situation is familiar to most
physicists. \ Boltzmann\cite{Boltzmann} assumed that thermal equilibrium
exists for noninteracting nonrelativistic particles in a uniform gravitational
field, or equivalently, in a uniformly accelerating box. \ Thermodynamic
arguments about cyclic lifting of a harmonic oscillator between the bottom and
top of the box indicate that the temperature must be uniform throughout the
box. \ Indeed, such arguments show that in equilibrium, the temperature is
constant throughout any nonrelativistic system. \ The temperature alone
determines the velocity distribution at any height. \ But then the velocity
distribution required to maintain the equilibrium spatial pressure gradient
against gravity or against acceleration is unique. \ If one now allows the
gravitational field or acceleration to go to zero, then one recovers the
Maxwell distribution for the equilibrium velocity distribution of particles in
thermal equilibrium in an inertial frame.\cite{Boltzmann}

In this article we wish to carry through an analogous argument for the
derivation of the equilibrium spectrum of classical relativistic scalar
radiation in an inertial frame. \ Unfortunately, the derivation is not as
simple as that for nonrelativistic particles because the radiation derivation
must use relativistic ideas, and these are not as familiar as those of
nonrelativistic mechanics. \ 

\subsection{The Rindler Frame}

Following the analogy with Boltzmann's work, we would like to discuss
radiation in a box undergoing uniform acceleration. \ Since we are dealing
with relativistic classical radiation, we would like to consider a box
undergoing uniform acceleration through Minkowski spacetime. \ In the frame of
the box, the acceleration should be constant in time, and the dimensions of
the box should not change so that the radiation pattern can be assumed steady
state. \ However, relativity introduces some complications which are quite
different from nonrelativistic kinematics. \ When viewed from an inertial
frame where the box is momentarily at rest at some $t=0$, the acceleration $a$
of a point of the box will appear to change according to the Lorentz
transformation for accelerations, $a=a^{\prime}/\gamma^{3}=a^{\prime}%
(1-v^{2}/c^{2})^{3/2}$, with the acceleration $a$ (seen in the inertial frame)
becoming smaller as the velocity $v$ of the box becomes larger even though the
acceleration $a^{\prime}$ in the frame of the box is constant in time.
\ Furthermore, in order for the box to maintain a constant length in its own
rest frame, the box must be found to undergo a length contraction in the
inertial frame. \ But this requires that different points of the box must
undergo different accelerations as seen in any inertial frame, and indeed, in
any inertial frame momentarily at rest with respect to the box. \ Thus the
proper acceleration of each point of the box must vary with height. \ This
relativistic situation has been explored in the literature\cite{Rindler} and
the coordinate frame associated with the box is termed a Rindler frame. \ If
the coordinates of an inertial frame are given by $ct,x,y,z,$ the coordinates
of the associated Rindler frame which is at rest with respect to the inertial
frame at $t=0$ are specified as\cite{RindlerF}
\begin{equation}
ct=\xi\sinh(\eta)
\end{equation}%
\begin{equation}
x=\xi\cosh(\eta)
\end{equation}
with $y$ and $z$ remaining unchanged between the frames and $\xi>0$. \ Using
these transformations, we see that the spacetime interval changes from the
Minkowski form in $(ct,x,y,z)$ over to a new form in the Rindler coordinates
($\eta,\xi,y,z)$%
\begin{align}
ds^{2}  &  =c^{2}dt^{2}-dx^{2}-dy^{2}-dz^{2}\nonumber\\
&  =\xi^{2}d\eta^{2}-d\xi^{2}-dy^{2}-dz^{2}%
\end{align}
If we consider a point with fixed spatial coordinates $\xi,y,z$ in a Rindler
frame, then (by introducing Eqs. (25) and (26) into the relation $\cosh
^{2}\eta-\sinh^{2}\eta=1)$ we find that in the inertial frame this point
follows the trajectory%
\begin{equation}
x=(\xi^{2}+c^{2}t^{2})^{1/2}%
\end{equation}
and has a constant proper acceleration given by%
\begin{equation}
a=c^{2}/\xi
\end{equation}
\ We notice from Eq. (29) that no single acceleration can be assigned to a
Rindler frame. \ Rather the acceleration varies with the coordinate $\xi,$
becoming very small for large $\xi$ and diverging as $\xi$ goes to zero. \ The
surface at $\xi=0$ (where the acceleration in Eq. (29) diverges) is termed the
"event horizon" of the Rindler frame. \ At any instant$\ $of Rindler time
$\eta,$ the spatial coordinates of the Rindler frame are in agreement with
those of a Minkowski frame which is instantaneously at rest with respect to
the Rindler frame.

\subsection{Variation in Temperature for Thermal Radiation in a Rindler Frame}

Although the temperature of thermal radiation is constant throughout
nonrelativistic systems in equilibrium, this constancy is not true in
relativistic gravitational physics, and, in particular, it is not true in a
Rindler frame. There are clearly profound differences between the
thermodynamics of nonrelativistic and relativistic physics. \ These profound
differences can be seen immediately in the contrasting determinations of the
forces $\mathbf{F}_{1}$ and $\mathbf{F}_{2}$ needed to accelerate from rest
respectively 1)a box of interacting particles and 2)a box of radiation. \ In
nonrelativistic physics, the force $\mathbf{F}_{1}$ accelerating the box of
particles is $\mathbf{F}_{1}=M\mathbf{a},$ where $M$ is the total mechanical
mass of the particles (independent of the potential energies in the box), and
the force $\mathbf{F}_{2}$ accelerating the box of radiation is zero
$\mathbf{F}_{2}=0$ since no mechanical mass is present. \ In relativistic
physics, both forces are given by $\mathbf{F}=(U/c^{2})\mathbf{a}$ where $U$
is the total energy in the box. \ In a Rindler frame, all unsupported systems
will tend to fall relative to the Rindler coordinates because the coordinates
of the frame are accelerating. \ Therefore for thermal equilibrium in this
relativistic system, the pressure (and hence the temperature) must increase at
lower depths in order to support the energy above it. \ (We can imagine
introducing massless horizontal reflecting surfaces into the box of radiation
and determining the pressure needed to support the thermal radiation above the
surface.) \ In relativity, the change of pressure $p$ with height due to
acceleration depends upon the sum $p+u$ of pressure plus thermal energy
density.\cite{temp} \ Since the thermal energy density $u$ depends on the
temperature as $T^{4}$, as shown in Eq,. (18), we expect
\begin{equation}
\frac{dp}{d\xi}=-\frac{p(T)+u(T)}{c^{2}}a=-\left(  \frac{1/3+1}{c^{2}}%
a_{Ss}T^{4}\right)  \left(  \frac{c^{2}}{\xi}\right)
\end{equation}
where we have use the connection of Eq. (19) $p=(1/3)u(T)=(1/3)a_{Ss}T^{4}.$
\ Thus the pressure at any point depends upon the temperature at that point
which depends in some unknown fashion upon the distance $\xi$ to the event
horizon. \ Therefore equation (30) becomes%
\begin{equation}
\frac{d}{d\xi}(\frac{1}{3}a_{Ss}T^{4})=\frac{4}{3}a_{Ss}T^{3}\frac{dT}{d\xi
}=-\frac{4}{3}a_{Ss}T^{4}\frac{1}{\xi}%
\end{equation}
or $dT/d\xi=-T/\xi.$ \ This equation has the solution $\ln(T)=-\ln
(\xi)+const,$ giving $T\xi=const$ or
\begin{equation}
T=const/\xi
\end{equation}
This result is consistent with the Tolman-Ehrenfest relation of general
relativity\cite{Tolman} $T(g_{00})^{1/2}=const$ for the change of temperature,
when we note from Eq. (27) that in the Rindler frame$\ (g_{00})^{1/2}=\xi.$
\ Thus we have found that there is no single temperature which can be assigned
to the thermal radiation in equilibrium in a Rindler frame. \ Rather the
temperature varies with the distance $\xi$\ from the event horizon, going to
zero at infinite distance and diverging on approach to the event horizon.

\subsection{Two-Point Correlation Function for Zero-Point Radiation in a
Rindler Frame}

Within classical physics, the zero-point radiation of the "vacuum state" is
present throughout spacetime and takes the same spectrum in any inertial
frame. \ This zero-point radiation will also be found in the Rindler frame
which is accelerating through Minkowski spacetime. \ Thus next we wish to
obtain the expression for the field correlation function for zero-point
radiation as evaluated in the Rindler frame. \ The correlation function can be
expressed in terms of the fields $\phi_{R}(\eta,\xi,y,z)$ seen in the Rindler
frame. \ The value of a scalar field at any spacetime point is independent of
the coordinate frame in which it is evaluated,
\begin{equation}
\phi_{R}(\eta,\xi,y,z)=\phi(ct,x,y,z)=\phi(\xi\sinh(\eta),\xi\cosh(\eta),y,z)
\end{equation}
Therefore merely substituting Eqs. (25) and (26) into Eq. (22), gives%
\begin{align}
\left\langle \phi_{R0}(\eta,\xi,y,z)\phi_{R0}(\eta^{\prime},\xi^{\prime
},y,z)\right\rangle  &  =\frac{-\hbar c}{\pi}[(\xi\sinh\eta-\xi^{\prime}%
\sinh\eta^{\prime})^{2}-(\xi\cosh\eta-\xi^{\prime}\cosh\eta^{\prime}%
)^{2}\nonumber\\
&  -(y-y^{\prime})^{2}-(z-z^{\prime})^{2}]^{-1}\nonumber\\
&  =\frac{-\hbar c}{\pi\lbrack2\xi\xi^{\prime}\cosh(\eta-\eta^{\prime}%
)-\xi^{2}-\xi^{\prime2}-(y-y^{\prime})^{2}-(z-z^{\prime})^{2}]}%
\end{align}
Although the correlation functions given in Eqs. (22) and (34) look quite
different, they actually involve the same zero-point radiation but described
in different coordinates. \ 

If we evaluate the zero-point correlation function at a single Rindler time
$\eta=\eta^{\prime},$ this corresponds to a single time $t=t^{\prime}$ in the
inertial frame which is momentarily at rest with respect to the Rindler frame.
\ In this case, the correlation functions in the inertial frame and in the
Rindler frame agree exactly, both involving the inverse square of the same
spatial distance between the field points. \ There is no spatial correlation
length appearing in the Rindler frame, just as there was none in the inertial
frame. \ The system is still a zero-point radiation system with no possibility
of defining a finite local energy density or local entropy density. \ However,
if we consider a single spatial point ($x,y,z)=$($x^{\prime},y^{\prime
},z^{\prime})$ in an inertial frame at two different times $t$ and $t^{\prime
},$ or a single spatial point ($\xi,y,z)=$($\xi^{\prime},y^{\prime},z^{\prime
})$ in the Rindler frame at two different times $\eta$ and $\eta^{\prime}$,
then the field correlation functions for zero-point radiation take quite
different forms. \ The correlation function for zero-point radiation in the
inertial frame is given in Eq. (23) while the expression in the Rindler frame
follows from Eq. (34) as
\begin{align}
\left\langle \phi_{R0}(\eta,\xi,y,z)\phi_{R0}(\eta^{\prime},\xi
,y,z)\right\rangle _{0}  &  =\frac{-\hbar c}{\pi\lbrack2\xi^{2}\cosh(\eta
-\eta^{\prime})-2\xi^{2}]}\nonumber\\
&  =\frac{-\hbar c}{\pi\lbrack2\xi\sinh\{(\eta-\eta^{\prime})/2\}]^{2}%
}\nonumber\\
&  =\frac{-\hbar c}{\pi\lbrack2\xi\sinh\{(\tau_{R}-\tau_{R}^{\prime}%
)/(2\xi/c)\}]^{2}}\nonumber\\
&  \frac{-\hbar ca^{2}}{\pi\lbrack2c^{2}\sinh\{(\tau_{R}-\tau_{R}^{\prime
})a/(2c)\}]^{2}}%
\end{align}
where in the third line we have introduced the proper time interval $(\tau
_{R}-\tau_{R}^{\prime})=\xi(\eta-\eta^{\prime})$ measured by a clock at rest
in the Rindler frame at the spatial coordinates $\xi,y,z,$ and in the fourth
line we have introduced the proper acceleration $a=c^{2}/\xi.$ \ In an
inertial frame, there are no characteristic lengths or times connected to any
coordinate point, and the field correlation function for zero-point radiation
given in Eq. (22)\ involves simply the Minkowski proper time interval between
any two spacetime points. \ However, each spatial coordinate point $\xi,y,z$
of the Rindler frame has associated a characteristic time $\xi/c$
(corresponding to the time for light to travel to the event horizon at speed
$c),$ and this is the characteristic time appearing in the third line of Eq.
(35) for the correlation function for zero-point radiation in Rindler
coordinates. \ The zero-point radiation in the Rindler frame is measured in
units of time which already contain a characteristic correlation time \ and
this correlation time is imposed on the zero-point correlation function given
in Eq. (35). \ 

\subsection{Example of a Horizontal Light-Clock in a Rindler Frame}

We wish to emphasize strongly that at every point of a Rindler frame, there is
a correlation time $\tau_{R}=\xi/c=c/a$ and an associated correlation
frequency $\Omega_{R}=1/\tau_{R}=a/c$ which is unrelated to the temperature of
any thermal radiation which may be present. \ Thus, for example, consider a
horizontal light clock of horizontal length $l$ at rest at height $\xi$ in a
Rindler frame. \ The Rindler coordinate time interval $\eta_{l}$ read by the
light clock corresponds to the time required for light to travel the
horizontal length $l.$ \ Now in the inertial frame which was momentarily at
rest with respect to the length $l$ when the light pulse started, the light is
seen to follow a diagonal path; this diagonal path has length $l$ in the
direction perpendicular to the Rindler-frame acceleration and a distance
$x(t)-x(0)=\xi\cosh\eta_{l}-\xi\cosh0$ in the direction parallel to the
Rindler-frame acceleration and occurs during an inertial-frame time interval
$t-0=(\xi/c)\sinh\eta_{l}-(\xi/c)\sinh0.$ \ Since in the inertial frame the
light travels with speed $c,$ we have $(ct)^{2}=[x(t)-x(0)]^{2}+l^{2}$. \ This
corresponds to%
\begin{equation}
(\xi\sinh\eta_{l})^{2}=(\xi\cosh\eta_{l}-\xi)^{2}+l^{2}%
\end{equation}
or%
\begin{align}
l &  =2\xi\sinh(\eta_{l}/2)\nonumber\\
&  =2\xi\sinh[\tau_{lR}/(2\xi/c)]\nonumber\\
&  =2\xi\sinh[\tau_{lR}(a/(2c)]
\end{align}
where $\tau_{lR}$ $=\xi\eta_{l}$ is the proper time read by a clock at height
$\xi$ in the Rindler frame. \ Thus the connection between the proper time
interval $\tau_{lR}$ and the length $l$\ of the horizontal light clock is
governed by the hyperbolic sine function as in Eq. (37). Accordingly we find
that for a small horizontal light clock in the Rindler frame, the time
interval $\tau_{lR}$ read by this horizontal light clock is given by the
linear relation $\tau_{lR}=\xi\eta_{l}=l/c,$ whereas for a large horizontal
light clock, there is an exponential connection between $l$ and $\tau_{lR}.$
\ The transition length between two these two regimes is given by the length
$l\approx\xi,$ the time $\tau_{lR}\approx\xi/c,$ and the associated frequency
$\Omega_{R}\approx c/\xi=a/c.$

\subsection{Assumption of a Single Correlation Time for Thermal Radiation in a
Rindler Frame}

At this point we want to consider the two-point field correlation function in
time for the radiation field $\phi_{RT}$ in the Rindler frame when thermal
radiation at temperature $T>0$ is present. \ We notice from Eq. (35) that in a
Rindler frame the two-point field correlation function for zero-point
radiation at a single spatial point already includes the finite correlation
time $\xi/c=c/a$ which is characteristic of the acceleration $a$ of a point
$\xi$ above the event horizon. \ Thus for small times $\tau_{R}$ (where the
high-frequency zero-point radiation contributes) the correlation function in
Eq. (35) behaves as $-\hbar c/(\pi c^{2}\tau_{R}^{2})$, whereas for long times
the behavior is as $-\hbar ca^{2}/\{\pi c^{4}\exp[\tau_{R}a/c]\}.$
\ Accordingly in a Rindler frame, both the acceleration $a$ and the finite
non-zero temperature $T$\ will contribute finite correlation times to the
two-field correlation function at fixed height $\xi$. \ Thus one might expect
three different time regions for the two-point field correlation function: i)
the short-time region dominated by high-frequency zero-point radiation, ii)
the region doinated by the acceleration-related correlation time, and iii) the
region dominated by the temperature-related correlation time. \ Depending upon
the relative magnitude of the acceleration $a$ and the temperature $T,$ the
two-point field correlation function and the associated frequency spectrum
would take on varying forms. \ This variation in form would allow us to
distinguish the relative temperature in the Rindler frame compared to the
acceleration, or the acceleration relative to the temperature. \ Since we have
seen in Eq. (32) that in a Rindler frame the temperature of the radiation must
behave with height $\xi$ as $T=const/\xi$ while the acceleration given in Eq.
(29) behaves with height $\xi$ as $a=c^{2}/\xi,$ the ratio of temperature $T$
to acceleration $a$ remains the same throughout the Rindler frame in thermal
equilibrium. \ We notice in Eq. (35) that in zero-point radiation the
correlation function in the Rindler frame is a monotonic function of the
acceleration $a$. \ Thermodynamics requires that the correlation function for
thermal radiation must also be a monotonic function of temperature $T.$
\ Clearly we want the field correlation function at two different times $\eta$
and $\eta^{\prime}$ when thermal radiation is present to fit with the
correlation function (35) when only zero-point radiation is present. \ The
simplest possibility is that the two-point correlation function of the fields
at a single height involves not two distinct correlation times but rather only
a single correlation time. \ This situation corresponds to substituting
$(a+const\times T)$ in place of the acceleration $a$ in the correlation
function of Eq. (35). \ This increase in the argument of the correlation
function corresponds to increasing the energy density in the normal modes
above the zero-point value, exactly as is appropriate for thermal radiation.
\ The only combination of fundamental units with the correct dimensions
requires the combination $a+\zeta2\pi ck_{B}T/\hbar$ where $\zeta$ is a
dimensionless number. \ The correlation function for thermal radiation in the
Rindler frame then takes the form%
\begin{equation}
\left\langle \phi_{RT}(\eta,\xi,y,z)\phi_{RT}(\eta^{\prime},\xi
,y,z)\right\rangle =\frac{-\hbar c}{\pi}\left(  \frac{(a+\zeta2\pi
ck_{B}T/\hbar)^{2}}{[2c^{2}\sinh\{(\tau_{R}-\tau_{R}^{\prime})(a+\zeta2\pi
ck_{B}T/\hbar)/(2c)\}]^{2}}\right)
\end{equation}
\ \ Furthermore, if we make this substitution, then we find that the
asymptotic limits are appropriate. \ We recall that at large values of $\xi,$
the acceleration of the Rindler frame becomes small so that the Rindler frame
has behavior similar to that of an inertial frame. \ But then in this
small-acceleration limit $a\rightarrow0$, we notice that the field correlation
function in time Eq. (38) for the Rindler frame takes just the same form as
the field correlation function in time Eq. (21) for an inertial frame; both
involve $T^{2}\mathfrak{F}$[$T(\tau-\tau^{\prime})].$ In the small temperature
limit $T\rightarrow0,$ the correlation function (38) returns to the zero-point
correlation function (35). \ At short time differences $(\tau_{R}-\tau
_{R}^{\prime}),$ the correlation function (38) still goes over to the
zero-point radiation result of Eq. (23) which is independent of both $a$ and
of $T.$ \ At long time differences $(\tau_{R}-\tau_{R}^{\prime}),$ the
correlation function decreases exponentially, but now with the combination
$a+\zeta2\pi ck_{B}T/\hbar.$ \ 

The correlation function in Eq. (38) corresponds to amplitudes for the normal
modes which have monotonically larger amplitudes with increasing temperature
(corresponding to increased energy due to the assignment of thermal energy)
than the amplitudes of the normal modes involving zero-point radiation alone.
\ At high temperature and fixed frequency, the thermal radiation dominates the
spectrum. \ On the other hand, in the limit as the temperature $T$ goes to
zero, the correlation function becomes the zero-point correlation function of
the Rindler frame given in Eq. (35). All of these considerations suggest that
the field correlation function given in Eq. (38) is indeed the thermal
correlation function for scalar radiation in a Rindler frame. \ In a Rindler
frame, thermal radiation is constrained to fit with zero-point radiation which
appeared from Lorentz invariance in a Minkowski frame. \ 

The appearance of only a single correlation time in the two-point field
correlation function at fixed height, as in Eq. (38), serves to hide the
acceleration of the system from any spatially-local measurement which
considers only time correlations. \ An observer who has access only to the
time correlation at a fixed spatial point would not be able to determine the
acceleration of the system since the correlation might represent any
combination of finite-temperature thermal radiation and acceleration through
zero-point radiation. \ However, measurements of spatial correlations in the
fields at fixed time will indeed allow separation of the acceleration and
finite-temperature aspects. \ In a sense, this behavior is analogous to the
suppression of acceleration information in a local measurement of particle
velocities in a nonrelativistic thermal distribution in an accelerating frame.
\ The velocity distribution at a fixed height in an accelerating frame is the
Maxwell distribution. \ The information about the acceleration of the frame is
contained in the spatial change in particle density with height.

\subsection{Planck Radiation Spectrum}

\ In the limit as the acceleration $a$ goes to zero while the temperature $T$
is held fixed, the proper time $\tau_{R}$ in the Rindler frame becomes equal
to the proper time $\tau_{M}$ in the Minkowski frame, and the field
correlation of the Rindler frame becomes that of thermal radiation in a
Minkowski frame. \ Indeed, if we consider the Minkowski frame limit
$a\rightarrow0$ in Eq. (38), then we find%
\begin{align}
\lim_{a\rightarrow0}\left\langle \phi_{RT}(\eta,\xi,y,z)\phi_{RT}(\eta
^{\prime},\xi,y,z)\right\rangle  &  =\left\langle \phi_{T}(ct,x,y,z)\phi
_{T}(ct^{\prime},x,y,z)\right\rangle \nonumber\\
&  =\frac{-\hbar c}{\pi}\left(  \frac{(\zeta2\pi ck_{B}T/\hbar)^{2}}%
{[2c^{2}\sinh\{(t-t^{\prime})(\zeta2\pi ck_{B}T/\hbar)/(2c)\}]^{2}}\right)
\end{align}
By taking the Fourier cosine transform of this correlation
function\cite{thermal}, we obtain the thermal radiation spectrum in a
Minkowski frame%
\begin{align}
\frac{\omega^{2}f^{2}(\omega)}{8\pi c^{2}} &  =\int_{0}^{\infty}d\tau
\frac{-\hbar c}{\pi}\left(  \frac{(\zeta2\pi ck_{B}T/\hbar)^{2}}{[2c^{2}%
\sinh\{\tau(\zeta2\pi ck_{B}T/\hbar)/(2c)\}]^{2}}\right)  \cos(\omega
\tau)\nonumber\\
&  =\frac{1}{2}\hbar\omega\coth\left(  \frac{\hbar\omega}{2\zeta k_{B}%
T}\right)
\end{align}
corresponding to an energy per normal mode from Eq. (6)%
\begin{equation}
U(\omega,T)=\frac{1}{2}\hbar\omega\coth\left(  \frac{\hbar\omega}{2\zeta
k_{B}T}\right)  =\frac{\hbar\omega}{\exp[\hbar\omega/(\zeta k_{B}T)]-1}%
+\frac{1}{2}\hbar\omega
\end{equation}
which is exactly the usual Planck scalar radiation result including zero-point
radiation\ when we set the unknown constant $\zeta=1$. \ At high frequencies
$\omega,$ the energy $U(\omega)$ becomes $U(\omega)=(1/2)\hbar\omega.$ \ At
low frequencies the energy $U(\omega)$ becomes $U(\omega)=k_{B}T.$ \ 

\section{Discussion}

The analysis given here has ties to work appearing in quantum field
theory.\cite{Crispino}\cite{Milonni} \ In connection with Hawking's ideas
regarding the quantum evaporation of black holes\cite{Hawking} and Fulling's
nonuniqueness of the field quantization,\cite{Fulling} Davies\cite{Davies} and
Unruh\cite{Unruh} noted the appearance of the Planck correlation function when
a point was accelerated through the quantum vacuum of Minkowski spacetime.
\ Within the quantum literature, a mechanical system accelerating through the
vacuum is often said to experience a thermal bath at temperature $T=\hbar
a/(2\pi ck_{B})$ and to take on a thermal distribution. \ There have been
controversies as to whether or not the acceleration turns the "virtual
photons" of the vacuum into "real photons." \ In this article, the analysis
has been entirely within classical physics. \ 

The analysis of thermal radiation given here is totally different from the
discussions which appear in text books of modern physics\cite{modes}. \ The
present analysis is crucially dependent upon relativistic physics whereas the
historical treatments combine nonrelativistic and relativistic aspects so that
the combination satisfies neither Galilean nor Lorentz invariance. The
emphasis upon a relativistic treatment in the present article is consistent
with recent analysis showing that scattering by relativistic (as opposed to
nonrelativistic) mechanical systems leaves the zero-point spectrum
invariant.\cite{invariant}

The classical analysis of thermodynamics in a Rindler frame has important
implications for the connections between classical and quantum theories which
will be pursued elsewhere.

\


\begin{thebibliography}{99}
\bibitem {corr1}See reference 7. \ The only difficult part of the integration
is the evaluation of the singular contribution where a temporary cut-off
$\exp[-v\Lambda]$ is introduced, and then the cut-off parameter $\Lambda$ is
taken to zero at the end of the calculation.
\end{thebibliography}
\end{document}